\documentclass[vecphys,amsmath,amssymb]{svmult}

\usepackage{multicol}        

\usepackage{makeidx}         
\usepackage{epsfig, psfrag, graphicx}        
\usepackage{multicol}        
\usepackage[bottom]{footmisc}
\usepackage{amssymb}

\newcommand{\be}{\begin{equation}}
\newcommand{\ee}{\end{equation}}
\newcommand{\ben}{\begin{eqnarray}}
\newcommand{\een}{\end{eqnarray}}
\newcommand{\bea}{\begin{eqnarray}}
\newcommand{\eea}{\end{eqnarray}}
\newcommand{\bdm}{\begin{displaymath}}
\newcommand{\edm}{\end{displaymath}}
\newcommand{\ba}{\begin{align}}
\newcommand{\ea}{\end{align}}
\newcommand{\lb}{\label}

\makeindex             

\begin{document}

\title*{Can the Arrow of Time be understood from Quantum Cosmology?}

\author{Claus Kiefer}

\institute{Institut f\"ur Theoretische Physik,
Universit\"at zu K\"oln, Z\"ulpicher
Stra\ss e 77, 50937 K\"oln, Germany.
\texttt{kiefer@thp.uni-koeln.de}
}

\maketitle


\begin{abstract}
I address the question whether the origin of the observed arrow of
time can be derived from quantum cosmology. After a general discussion of
entropy in cosmology and some numerical estimates, I give a brief
introduction into quantum geometrodynamics and argue that this
may provide a sufficient framework for studying this question. I then
show that a natural boundary condition of
low initial entropy can be imposed on the universal wave function. The
arrow of time is then 
correlated with the size of the 
Universe and emerges from an increasing amount of decoherence due to
entanglement with unobserved
degrees of freedom. Remarks are also made concerning the arrow of time in 
multiverse pictures and scenarios motivated by dark energy.       
\end{abstract}

\noindent To appear in {\em The Arrow of Time}, edited by
    L.~ Mersini-Houghton and R.~Vaas (Springer, Berlin, 2010).

\section{ Introduction}

The fundamental laws of physics, as they are presently known, are 
mostly invariant with respect to a reversal of time: to every solution
there exists an equally viable solution in which $t$ is replaced by
$-t$. The only exceptions 
are some processes described by the weak interaction, but these cases
can also be subsumed under time-reversal invariance in the broad sense
because its violation there can directly be compensated by an application of a
unitary CP-transformation. 
  
Despite this fundamental invariance, most classes of phenomena
observed in Nature distinguish a specific direction of time. 
These are the famous ``arrows of time'' which are discussed at length
in \cite{Zeh1}, see also \cite{Zeh2} and other contributions to this
volume. This observed discrepancy between the symmetric laws and the
asymmetric facts does not constitute an inconsistency; asymmetric
phenomena are compatible with symmetric laws, and most solutions of
the fundamental equations are not symmetric by themselves. What is
peculiar is the fact that the time direction of the phenomena is always
the same. 

Usually one distinguishes between various manifestations of the arrow
of time \cite{Zeh1}. The electrodynamic arrow expresses the preference
of the retarded over the advanced solutions. The thermodynamic arrow
is given by the non-decrease of entropy for closed systems, as
expressed by the Second Law of thermodynamics. The electrodynamic
arrow can, in fact, be derived from the Second Law by using the
thermodynamical properties of absorbers. The quantum-mechanical arrow
expresses a direction through the measurement process or, in the
Everett picture, the branching of the wave function. A central role is
there played by decoherence -- the irreversible and unavoidable
emergence of classical properties through interaction with the
environment. Finally, gravitational systems exhibit a preferred
direction either through gravitational collapse or through the
expansion of the Universe. The question raised by the presence of all
these arrows is whether a common {\em master } arrow of time is behind
all of them.

One might wonder whether the arrow of time points into different directions for
different subsystems of the Universe, for example, for different galaxy
clusters. Using arguments that can be traced back to
Emile Borel in the 1920s, one can recognize that there is no strict isolation
of subsystems and that therefore all arrows in the Universe must point into the
{\em same} direction. This suggests that the master arrow of time may
be found in cosmology.

Can one explain the presence of these distinguished time directions in
the framework of physics? Is there a master arrow of time and, if yes,
where does it come from? One might speculate that a new, hitherto
unknown, fundamental law exists, which is asymmetric in time. Models
of wave-function collapse are explicit examples for such new
laws. However, no empirical evidence exists for them. The
alternative to such a speculation is the presence of a distinguished
cosmic boundary condition of low entropy at or near the Big Bang. One
would then expect that the entropy of the Universe increases in the
direction of increasing cosmic size. The question remains, however, where such a
boundary condition could come from. 

As indicated by the singularity theorems of general relativity, a
consistent description of the Big Bang may require a new framework such
as quantum gravity. The question then arises whether the origin of the
arrow of time can be understood there. This is the topic
of my essay. I shall start in the next section by making more precise
the arguments in favour of a cosmic boundary condition of low
entropy. I shall then present a framework of quantum gravity in which
the above question can be addressed -- quantum geometrodynamics, the
direct quantization of Einstein's theory of general relativity. In the
last section I shall then present how, in fact, the origin of
irreversibility could be understood from quantum cosmology. I shall
also speculate there about possible quantum effects and the fate of
the Second Law in the future of our Universe. The Appendix contains
numerical estimates concerning the entropy and the maximal entropy of
our Universe. 

\section{Entropy and Cosmology}

Already Ludwig Boltzmann had speculated that the Second Law has its
origin in a gigantic fluctuation in the Universe. His picture was that
the Universe is eternally existing and at its maximal entropy most of
the time, but 
that at very rare occasions (which, of course, can happen in an
eternal Universe) the entropy fluctuates to a very low value from
which it will then increase; this would then enable our existence and
lead to the arrow of time that we observe. The weak point in this
argument was disclosed in the 1930s by Carl Friedrich von
Weizs\"acker: if one takes into account the possibility of entropy
fluctuations, a fluctuation that produces at once the world that we
observe including our existence and our memories, although by itself
extremely unlikely, would still be much more probable than Boltzmann's
fluctuation which has to create the whole history of the world in
addition to the present state. Strange observers which according
to such a fluctuation could spontaneously pop into existence have
recently been discussed in the context of a ``multiverse'' picture and
there been called ``Boltzmann brains'', cf. \cite{Bbrains} and the
references therein. The multiverse picture can be motivated by
inflationary scenarios of the early Universe 
(e.g. Linde's ``eternal inflation'') and describes the full
Universe as being infinitely extended and very inhomogeneous on large
scales, but containing many Friedmann subuniverses of the kind that we
observe. In such a gigantic Universe, even the tiniest entropic
fluctuation would occur somewhere. I shall briefly address the
multiverse picture below, but focus in the following on the observable
part of the Universe, which is approximately homogeneous and
isotropic.

In order to know how special our Universe in fact is, one would like
to calculate both the actual entropy of our Universe as well as the
maximal possible entropy. The non-gravitational entropy is dominated
by the photons of the Cosmic Microwave Background (CMB) radiation; it
contributes about $2\times 10^{89}\ k_{\rm B}$ \cite{EL09}. 
Linde and Vanchurin have, moreover, given an estimate of an upper
limit for the non-gravitational entropy, which would be obtained if
all particles were ultra-relativistic: their value is about 
$10^{90}\ k_{\rm B}$ and thus only about one order of magnitude more
than the CMB value \cite{LV09}.
These are 
very large numbers, but they are much smaller than the gravitational
contribution to the entropy. 
Unfortunately, a general expression for
gravitational entropy does not exist. Because of the universal
attractivity of gravity, one can only expect that gravitational
entropy increases during a gravitational collapse, in contrast to the
entropic trend of ordinary matter which prefers a homogeneous state. 
However, for the most extreme case of gravitational collapse, an
entropy formula exists: the Bekenstein--Hawking entropy for black
holes. It reads
\begin{equation}
\label{SBH}
S_{\rm BH}=\frac{k_{\rm B}c^3A}{4G\hbar}=k_{\rm B}\frac{A}{4l_{\rm P}^2}
 \ ,
\end{equation}
where $A$ is the surface area of the event horizon and $l_{\rm
  P}=\sqrt{G\hbar/c^3}$ is
the Planck length; in the following estimates we shall set Boltzmann's
constant $k_{\rm B}$ equal to one.

To see how large the Bekenstein--Hawking entropy can become, let us
estimate its value for the Galactic Black Hole -- the supermassive
black hole in the centre of our Milky Way with a mass
$M\approx 3.9\times 10^6\ M_{\odot}$.
Neclecting its angular
momentum, which would anyway decrease the estimated entropy, one gets
from (\ref{SBH})
\be
S_{\rm GBH}=\pi\left(\frac{R_{\rm S}}{l_{\rm P}}\right)^2
\approx 6.7\times 10^{90}\ ,
\ee
where $R_{\rm S}$ denotes the Schwarzschild radius. This already
exceeds by more than one order of magnitude
the non-gravitational contribution to the entropy. According to a recent
investigation \cite{EL09}, all supermassive black holes together yield
an entropy of $S=3.1^{+3.0}_{-1.7}\times 10^{104}$.

Roger Penrose has pointed out in \cite{Penrose1} that the maximal
entropy for the observable Universe would be obtained if all its
matter were assembled into one black hole. Taking the most recent
observational data, this would yield the entropy (calculated in the
Appendix) 
\be
\lb{Smaxnew}
S_{\rm max}\approx 1.8\times 10^{121} \ . 
\ee
This may not yet be the maximal possible entropy. Our Universe
exhibits currently an acceleration which could be caused by a
cosmological constant~$\Lambda$. If this is true, it will expand
forever, and the entropy in the far future will be dominated by the
entropy of the cosmological event horizon
(the ``Gibbons--Hawking entropy'' \cite{GH77}). The estimate, presented in
the Appendix, yields
\be
\lb{Sgh}
S_{\rm GH}=\frac{3\pi}{\Lambda l_{\rm P}^2}\approx 2.88\times
10^{122}\ ,
\ee
which is about one order of magnitude higher than (\ref{Smaxnew}).

Following the arguments in \cite{Penrose1}, the ``probability'' for
our Universe can be estimated as
\be
\frac{\exp(S)}{\exp(S_{\rm max})}\approx \frac{\exp(3.1\times
  10^{104})}{\exp(2.9\times 10^{122})}\approx \exp(-2.9\times
10^{122})\ .
\ee
Our Universe is thus very special indeed. It must have ``started''
near the Big Bang with an extremely low entropy; the Universe must
have been very smooth in the past, with no white holes being present. 
Penrose has reformulated this observation in his Weyl-tensor hypothesis:  
the Weyl tensor is zero near the Big Bang (describing its smooth
state), but diverges in a Big Crunch (provided the Universe will
recollapse). Since the Weyl tensor describes in particular
gravitational waves, this hypothesis entails that all gravitational
waves must be retarded. This is analogous to the Sommerfeld condition
in electrodynamics and the absence of advanced electromagnetic waves
\cite{Zeh1}. There, the electromagnetic arrow can be traced back to
the thermodynamic arrow and the Second law by using the thermodynamic
properties of absorbers, but this is not possible here because
gravitational waves are too weak. 

The Weyl-tensor hypothesis is not yet an explanation, but only a
description of the low initial cosmic entropy. Penrose has recently
reformulated his hypothesis in the framework of his ``Conformal
Cyclic Cosmology'' (CCC) \cite{Penrose2}. 
A central role is attributed therein to a
proposed information loss in black holes and the ensuing nonunitary
evolution. His whole picture, however, remains classical as far as
gravity is concerned. Here, I would like to adopt instead the point of view
that the gravitational field is fundamentally of quantum
nature. This is not a logical necessity, but one can put forward
physical arguments in favour of quantum gravity as the more
fundamental theory \cite{OUP,AKR08}. Although
there exist non-singular classical solutions, the singularity theorems
of classical relativity suggest the abundance of singularities in the
classical theory; a more general framework is therefore needed to
exorcize them. Moreover, gravity acts
universally, so it couples to all other fields of Nature, all of which
are so far described by quantum theory. It could then at least be considered
unnatural to leave the gravitational field classical; this would
become especially awkward in the context of a unified theory of all
interactions. 

In the next section I shall briefly describe an approach to quantum
gravity which is very conservative and which despite its limits should
be able to provide insights into the origin of the arrow of time.

\section{Quantum Geometrodynamics}

A full quantum theory of gravity remains elusive \cite{OUP}. Can one
nevertheless say something reliable about quantum gravity
without knowing the exact theory? In \cite{GRG} I have made the point
that this is indeed possible. The situation is analogous to the role
of the quantum mechanical Schr\"odinger equation. Although this equation
is not fundamental (it is non-relativistic, it is not
field-theoretic), important insights can be drawn from it. For
example, in the case of the hydrogen atom, one has to impose boundary
conditions for the wave function at the origin $r\to 0$, that is, at
the centre of the atom. This is certainly not a region where one would
expect non-relativistic quantum mechanics to be exactly valid, but its
consequences, in particular the resulting spectrum, are empirically
correct to an excellent approximation. 

Erwin Schr\"odinger has found his equation by ``guessing'' a wave equation
from which the Hamilton--Jacobi equation of classical mechanics can be
recovered in the limit of small wavelengths, analogously to the limit
of geometric optics from wave optics. The same approach can be applied
to general relativity. One can start from the Hamilton--Jacobi version
of Einstein's equations and ``guess'' a wave equation from which they
can be recovered in the classical limit. The only assumption that is
required is the universal validity of quantum theory, that is, its
linear structure. It is not yet needed for this step to impose a
Hilbert-space structure. Such a structure is employed in quantum
mechanics because of the probability interpretation for which one
needs a scalar product and its conservation in time (unitarity). The
status of this interpretation in quantum gravity remains open, see
below. 

The result of this approach is quantum geometrodynamics. Its central
equation is the Wheeler--DeWitt equation, first discussed by Bryce
DeWitt and John Wheeler in the 1960s. In a short notation, it is of
the form
\be
\lb{WdW}
\hat{H}\Psi=0\ ,
\ee
where $\hat{H}$ denotes the full Hamiltonian for both the gravitational
field (here described by the three-metric) as well as all non-gravitational
fields. For the detailed structure of this equation I refer, for
example, to the classic paper by DeWitt \cite{DeWitt} or the general
review in \cite{OUP}. Two properties are especially important for
our purpose here. First, this equation does not contain any classical
time parameter $t$. The reason is that spacetime as such has
disappeared in the same way as particle trajectories have disappeared
in quantum mechanics; here, only space (the three-geometry)
remains. Second, inspection of $\hat{H}$ exhibits the local hyperbolic
structure of the Hamiltonian, that is, the Wheeler--DeWitt equation
possesses locally the structure of a Klein--Gordon equation. In the
vicinity of Friedmann universes, this hyperbolic structure is not only
locally present, but also globally. One can thus define a new time
variable which exists only intrinsically and which can be constructed
from the three-metric (and non-gravitational fields) itself. It is
this absence of external time that could render the probability
interpretation and the ensuing Hilbert-space structure obsolete in
quantum gravity, for no conservation of probability may be 
needed.\footnote{The situation is different for an isolated quantum
  gravitational system such as a black hole; there, the semiclassical
  time of the rest of the Universe enters the description \cite{KMM09}.}

How, then, can one understand the emergence of an arrow of time from a
fundamental equation which is itself timeless? I shall address this
issue in the next section.

\section{Arrow of Time from Quantum Cosmology}

Quantum cosmology is the application of quantum theory to the Universe
as a whole. In a first approximation, the Universe is homogeneous and
isotropic. The three-metric is then fully characterized by the scale
factor, $a$, of the Universe. Classically, the Universe evolves in
time, $a(t)$; the same holds for the matter fields. In quantum
cosmology, $t$ has disappeared and all available information is
encoded in the wave function $\psi(a,\ldots)$, where the $\ldots$
denote homegeneous matter degrees of freedom. For the simple
two-dimensional configuration space consisting of the scale factor and
a minimally coupled scalar field $\phi$, the Wheeler--DeWitt equation
reads (with $c=1$)
\bea
\lb{whdw1}
\hat{H}\Psi&=&\left(
\frac{2\pi G\hbar^2}{3}\frac{\partial^2}{\partial\alpha^2}-
\frac{\hbar^2}{2}\frac{\partial^2}{\partial\phi^2}\right.\nonumber\\ & & +
\left.e^{6\alpha}\left(V\left(\phi\right)+\frac{\Lambda}{8\pi G}\right)
-3e^{4\alpha}\frac{k}{8\pi G}\right)\Psi(\alpha,\phi)=0\ ,
\eea
with cosmological constant $\Lambda$ and curvature index $k=\pm 1,
0$. The variable $\alpha=\ln a$ has been introduced for convenience. 

In order to discuss thermodynamical issues, additional degrees of freedom
must be added. One option is to consider small inhomogeneities in the
vicinity of homogeneity. This can be achieved, for example, through a
multipole expansion on the three-sphere (assuming the Universe is
closed) \cite{HH85,CK87}. Schematically, the Wheeler--DeWitt equation
(\ref{WdW}) then assumes the form
\begin{equation}
\lb{WdW2}
 \hat{H} \, \Psi = \left(\frac{2\pi G\hbar^2}{3}
\frac{\partial^2}{\partial\alpha^2} + \sum_i \, \left[
-\frac{\hbar^2}{2}
\frac{\partial^2}{\partial x_i^2}+\underbrace{V_i(\alpha,x_i)}_{\to 0\ 
{\rm for}\ \alpha
\rightarrow -\infty}\right]\right) \, \Psi = 0 \ ,
\end{equation}
where the $\{ x_i\}$ denote the scalar field as well as the inhomogeneous
degrees of freedom;
$V_i(\alpha,x_i)$ are the corresponding potentials. One
recognizes immediately that this Wheeler--DeWitt equation is
hyperbolic with respect to the intrinsic time $\alpha$. Initial
conditions are thus most naturally formulated with respect to constant
$\alpha$. 

The important observation is now that the potential in (\ref{WdW2}) is
{\em asymmetric} with respect to $\alpha$; if written out, it contains
explicit factors of $\E^{6\alpha}$, etc., and vanishes in the limit
$\alpha\to-\infty$.  In contrast to almost all
the other fundamental equations in physics, it thereby distinguishes a
direction in (intrinsic) time. 
One could thus envisage of solution to the Wheeler--DeWitt equation,
which near the Big Bang would be an approximate product state between
all degrees of freedom \cite{Zeh1},
\begin{equation}
\lb{psiBB}
\Psi \quad \stackrel{\alpha \, \to \, -\infty}{\longrightarrow}\
\psi_0(\alpha)\prod_i \psi_i(x_i)\ .
\end{equation}
Introducing the entropy of the Universe as an entanglement entropy, in
which irrelevant, that is, unobservable or unobserved degrees of
freedom (such as small gravitational waves 
described by some of the $x_i$) are integrated out, the state
(\ref{psiBB}), which is a product state, would yield a vanishing
entropy. For increasing $\alpha$, this solution would evolve into a
superposition of $\alpha$ and the inhomogeneous modes (as well as
between the inhomogeneous modes). Integrating out all or part of the
$x_i$ would then yield a non-vanishing and increasing entropy with respect to
$\alpha$. Increasing entanglement would then cause increasing
decoherence for the 
relevant degrees of freedom \cite{Zeh3,CK87,deco}. Decoherence -- the
unavoidable emergence of classical properties through interaction with
irrelevant degrees of freedom -- is perhaps the most fundamental
irreversible process and thus stands behind all arrows of time
\cite{Zeh1,deco}. Because of the asymmetry of (\ref{WdW2}) with
respect to $\alpha$, substituting $\alpha\to -\alpha$ would {\em not}
yield a solution of the Wheeler--DeWitt equation. If a solution of the
form (\ref{WdW2}) were the one describing our Universe, we could
understand from it the irreversible appearance of our world. A full
understanding of quantum gravity would perhaps single out a unique
solution for the Wheeler--DeWitt equation, a possibility already
envisioned by DeWitt \cite{DeWitt}.

There are indications that the above quantum state would
evolve into a symmetric state where all perturbations are in an
(at least approximate)
de~Sitter-invariant vacuum state. Such a vacuum state is a good
candidate for the early Universe \cite{Starobinsky}.
The state for each perturbation mode would describe a superposition of
inhomogeneous states, that is, a non-classical state. However,
the mechanism of decoherence also comes into play here, generating a
classical behaviour for the modes which may then serve as the seeds
for the origin
of structure in the Universe \cite{pointerstates}.

We have not yet discussed the connection between the intrinsic time
$\alpha$ and the ``observed time'' $t$ which should be at our disposal
at least in an appropriate semiclassical limit. This can be achieved
through a Born--Oppenheimer type of approximation scheme,
cf. \cite{OUP} and the references therein. Some degrees of freedom
such as the scale factor may serve as the semiclassical variables from
which a semiclassical time $t$ can be defined in appropriate
situations. This will be the time variable which controls the dynamics
in this approximation and which enters an effective and approximate
Schr\"odinger equation for the non-gravitational quantum
variables. The arrow of time aligned along increasing scale factor $a$
thus trivially extends to the semiclassical time $t$ -- as long as the
semiclassical approximation is valid.

The above ideas may, with slight elaborations, also apply to the idea
of the multiverse, that is, to a Universe with many approximately
homogeneous sub-universes, cf. \cite{LV09} and the references
therein. Quantum entanglement is not limited to sub-horizon scales and
may thus be effective also in the full multiverse. Decoherence should
then distinguish the same arrow of time everywhere in the multiverse.
Applying also here the idea that quantum fluctuations, after their
effective classicality due to decoherence, become the seeds for galaxy
formation, Linde and Vanchurin estimate in \cite{LV09} the number of
realizations of the emergent classical fluctuations in the gigantic
multiverse. This
number would also correspond to the number of branches of the universal
wave function in the Everett interpretation when applied to our Hubble domain.  
After decoherence, each realization can serve as a classical initial
condition for the subsequent evolution of the Universe. They find for
the total number of distinguishable locally ``Friedmann universes'' the
estimate 
\be
\E^{S_{\rm pert}}\lesssim \E^{\E^{3N}}\ ,
\ee
where $S_{\rm pert}$ is the total entropy of the perturbations, see
also \cite{pointerstates,verbier}, and $N$ is the number of e-folds of
slow-roll (post-eternal) inflation. In the simplest models of chaotic
inflation, one thereby gets the incredibely high number \cite{LV09}
\bdm
10^{10^{10^7}} \ .
\edm  
(A much lower number -- two instead of three exponentials -- is
obtained in the case of a positive cosmological constant.) Adopting,
in addition, the landscape picture of string theory, this estimate
would correspond to the case of one vacuum. Taking all the vacua into
account, the number will be even higher. The issue of the
Wheeler--DeWitt equation on a configuration space mimicking the
landscape picture and the question of a low-entropy initial condition
was discussed, for example, in \cite{HMH}.

The idea of quantum cosmology is that the whole Universe at all scales
is described by quantum theory. Thus, a priori, quantum effects are
not restricted to the Planck scale. In the case of a classically
recollapsing quantum universe, for example, one can predict the
occurrence of quantum effects near the classical turning point
\cite{KZ}, see also \cite{Zeh1,KS09} for a detailed
explanation. Because the arrow of time in the above scenario is
correlated with increasing scale factor $a$, the quantum universe would in
this case consist of many branches in which the arrow of time always
points in the direction of increasing $a$. These branches would be
decohered components of the universal wave functions
and would thus become independent of each other for most of their
existence, but they would interfere destructively at the classical
turning point in order to fulfill the boundary condition $\Psi\to 0$
for $a\to\infty$, which is necessary for a model in which the classical
trajectories recollapse. Consequently, no classical observers would be
able to survive a transition through the turning point, and time as
well as the classical evolution would come to an end there. 

This is an impressive example for a quantum effect at large scales. 
Other examples can be found in models that are of interest because
they may describe a dynamical dark energy in our Universe. Examples
are models which classically exhibit a big-rip or a big-brake
singularity; in the first case, the Universe can become infinitely
large in a finite time, while in the second case it comes 
to an abrupt halt in the future. In both cases, this corresponds to a
singular region. One can now study solutions of the corresponding
Wheeler--DeWitt equations and finds that the singularities will be
avoided, cf. \cite{KS09} and the references therein: the semiclassical
approximation breaks down when approaching the region of the classical
singularity, and for the big brake the wave function even becomes zero
there. What are the consequences of this scenario for the arrow of
time? 

Since the semclassical time comes to an end, so does the arrow.
The Universe enters a genuine quantum era which no classical observers
(and others are not known)
could survive. This is analogous to the above discussed turning
point. The world then becomes truly timeless.

One might wonder what happens in the case of models which classically
describe bouncing cosmologies \cite{bounces}: the Universe would then
undergo many, perhaps infinite, cycles of expansion and
recollapse. What would happen with the entropy in these cases? If the
entropy were indeed correlated with the scale factor, as the scenario discussed
above suggests, the arrow of time would not continue through a turning
point. The bouncing models would thus make no sense in quantum
cosmology; one would only have branches of the wave function in which
the arrow would point from small to large universe and where time
would end when approaching a classical turning point. 

We have restricted the discussion to quantum geometrodynamics. At
least for scales above the Planck length, which includes the above discussed
quantum scenarios for a big universe in the future, this should
provide a reliable framework. Modifications are, however, expected
when approaching the Planck regime, that is, the region of the Big
Bang. Such modifications have been addressed in string theory and loop
quantum gravity \cite{OUP}. In the case of loop quantum cosmology, the
Wheeler--DeWitt equation is replaced at the most fundamental level
by a difference (instead of differential) equation. For a big
universe, the differences 
to quantum geometrodynamics are negligible; this concerns, for
example, the examples of the big rip and the big brake.
Near the big-bang singularity, however, the situation is different. 
The emerging scenario is discussed at length in another contribution
to this volume \cite{Bojowald}. Also there, the author suggests ``the
possibility of deriving a beginning within a beginningless theory''. 
Thus, although the approaches may be different, the common fundamental
challenge is to understand the observed time and its arrow from a
scenario of the world which is fundamentally timeless.

\section*{Acknowledgements}

I thank Max D\"orner and Tobias Guggenmoser for a
careful reading of this manuscript.  

\begin{appendix}
\section{Some Numerical Estimates}
\lb{appendix}

Here we recapitulate the numerical estimates about the maximal
possible entropy in the Universe, presented by Penrose in
\cite{Penrose1}, in the light of recent cosmological data
\cite{Hinshaw}. Since our Universe is spatially flat to a high degree
of accuracy, the mass of the matter (both visible and dark) in our
present Hubble volume is given by
\be
\lb{MU}
M_{\rm U}=\frac{4\pi\rho_{\rm m}c^3}{3H_0^3}\ ,
\ee
where $\rho_{\rm m}$ is the matter density, and $H_0\approx
70.5\ \mathrm{km/s\ Mpc}\approx 2.27\times 10^{-18}\ \mathrm{s}^{-1}$
is the Hubble constant. Introducing the critical density
\be
\rho_{\rm c}=\frac{3H_0^2}{8\pi G}\ ,
\ee
we can use the density parameter $\Omega_{\rm m}= \rho_{\rm
  m}/\rho_{\rm c}\approx 0.274$ and write
\be
M_{\rm U}=\frac{c^3\Omega_{\rm m}}{2GH_0}\ .
\ee
In order to estimate the maximal entropy, we shall assume that the Universe
up to the Hubble scale consists of one black hole with mass $M_{\rm U}$. 
Since our present Universe is dominated
by dark energy, which for our purpose here can be approximated by a
cosmological constant $\Lambda$, we have to take into account that the
metric outside this hole is, in fact, the Schwarzschild--de~Sitter
metric, see, for example, \cite{Geyer}.  
Numerically we have $\Lambda\approx
1.25\times 10^{-56}\ \mathrm{cm}^{-2}$ and $\Omega_{\Lambda}=\Lambda
c^2/3H_0^2\approx 0.726$ \cite{Hinshaw}. In the
Schwarzschild--de~Sitter metric, the black hole has a maximal mass
given by
\be
\lb{MN}
M_{\rm N}=\frac{c^3}{3\sqrt{3\Omega_{\Lambda}}GH_0}\approx 
 4\times 10^{55}\ {\rm g}\ ,
\ee
which corresponds to the case of the Nariai metric (therefore the
index N). We thus have to check whether $M_{\rm U}$ is greater or
smaller than $M_{\rm N}$; only in the latter case can the Universe accommodate
one single black hole. A short calculation shows
\be
\frac{M_{\rm U}}{M_{\rm N}}=\frac{3\sqrt{3\Omega_{\Lambda}}\Omega_{\rm
    m}}{2} \approx 0.61\ ,
\ee
so all of $M_{\rm U}$ can indeed be assembled into one black hole.  

We now assume that the maximal entropy is given by a Schwarzschild
black hole with mass $M_{\rm U}$ (a non-vanishing angular momentum
would give a smaller entropy). Thus,
\be
S_{\rm max}=\pi\left(\frac{R_{\rm h}}{l_{\rm P}}\right)^2\ ,
\ee
where $R_{\rm h}$ denotes the radius of the black-hole event horizon
(as opposed to the cosmological horizon $R_{\rm c}$). In the
Schwarzschild--de~Sitter metric we have
\be
R_{\rm h} =
\frac{3GM_{\rm U}\ell\xi}{c^2}\left(1-\sqrt{1-\frac{1}{\ell\xi^3}}\right)
=\frac{\xi}{\sqrt{\Lambda}}\left(1-\sqrt{1-\frac{1}{\ell\xi^3}}\right)\ ,
\ee
where $\ell^{-1}=M_{\rm U}/M_{\rm N}$ and
$\xi=\cos(\frac13\cos^{-1}[\ell^{-1}])
\approx 0.95$.
With the above numbers we have 
\be
R_{\rm h}\approx 2.13 \frac{GM_{\rm U}}{c^2}\approx 0.38\times
10^{28}\ {\rm cm}
\ee
and therefore
\be
\lb{Smax}
S_{\rm max}\approx 1.8\times 10^{121} \ . 
\ee
This is the number that should replace the estimate $10^{123}$ in
\cite{Penrose1} if one makes use of the data presented in
\cite{Hinshaw}. 

Expressed in grams, the mass (\ref{MU}) is $M_{\rm U}\approx
2.4\times10^{55}\ \mathrm{g}$ and therefore corresponds to about
$1.5\times 10^{79}$ baryons. In the case of $10^{80}$ baryons, as used
in \cite{Penrose2}, one would find $\ell^{-1}>1$, that is, the
corresponding mass would exceed the Nariai mass (\ref{MN}) and it
would thus not be possible to assemble this mass into a single black
hole.

In the case of a non-vanishing cosmological constant there occurs also
the gravitational entropy $S_{\Lambda}$ associated with the
cosmological event horizon $R_{\rm c}$ \cite{GH77}, where
\be
R_{\rm c} =
\frac{3GM_{\rm U}\ell\xi}{c^2}\left(1+\sqrt{1-\frac{1}{\ell\xi^3}}\right)
\approx 1.29\times 10^{28}\ {\rm cm}\ .
\ee
It reads
\be
S_{\Lambda}=\pi\left(\frac{R_{\rm c}}{l_{\rm P}}\right)^2
\approx 1.99\times 10^{122}\ .
\ee
One should thus in principle consider the sum of $S_{\Lambda}$ and the
entropy associated with all matter being trapped in a single black hole.  
However, the maximal entropy is reached for asymptotic times
$t\to\infty$ when the matter content becomes irrelevant 
(because it will be diluted and no black hole with mass $M_{\rm U}$
will be formed); the radius of the event horizon then approaches
\be
R_{\rm c}=\sqrt{\frac{3}{\Lambda}}\approx 1.55\times 10^{28}\ {\rm
  cm}\ .
\ee
The entropy associated with this event horizon then approaches the
``Gibbons--Hawking'' entropy \cite{GH77}
\be
\lb{SGH}
S_{\rm GH}=\frac{3\pi}{\Lambda l_{\rm P}^2}\approx 2.88\times
10^{122}\ ,
\ee
which is about $16$ times the black-hole maximal entropy
(\ref{Smax}). The numerical value in (\ref{SGH}) is also
presented in \cite{EL09}. 

Taking the case of the Nariai mass (\ref{MN}), one would have the
total entropy $S_{\rm N}+S_{\Lambda}=2S_{\rm GH}/3$, which would give
further support to consider (\ref{SGH}) as the maximal possible
entropy of the observable Universe, as suggested by current observational data.

\end{appendix}




\begin{thebibliography}{99.}

\bibitem{Zeh1} H. D. Zeh (2007): {\em The physical basis of the
    direction of time}, fifth edition (Springer, Berlin).
\bibitem{Zeh2} H. D. Zeh (2009): Open questions regarding the arrow of
  time. Contribution to this volume, see also
  arXiv:0908.3780v2 [gr-qc].
\bibitem{Bbrains} A. De Simone, A. H. Guth, A. Linde, M. Noorbala,
  M. P. Salem, and A. Vilenkin (2008): Boltzmann brains and the
  scale-factor cutoff measure of the multiverse. arXiv:0808.3778v1
  [hep-th].
\bibitem{EL09} C. A. Egan and C. H. Lineweaver (2009): A larger
  estimate of the entropy of the universe. arXiv:0909.3983v1
  [astro-ph.CO].
 \bibitem{LV09} A. Linde and V. Vanchurin (2009): How many universes
  are in the multiverse? arXiv:0910.1589v1 [hep-th]. 
\bibitem{Penrose1} R. Penrose (1981): Time-asymmetry and quantum gravity.
           In {\em Quantum gravity}, Vol.~2, edited by C. J. Isham,
           R.~Penrose, and D. W.~Sciama,~pp.~242--272 (Clarendon
           Press, Oxford).
\bibitem{GH77} G. W. Gibbons and S. W. Hawking (1977): Cosmological
  event horizons, thermodynamics, and particle creation. {\em
    Phys. Rev. D} {\bf 15}, 2738--2751. 
\bibitem{Penrose2} R. Penrose (2009): Black holes, quantum theory and
  cosmology. {\em J. Phys.: Conference Series} {\bf 174}, 012001.
\bibitem{OUP} C. Kiefer (2007): {\em Quantum Gravity},
second edition (Oxford University Press, Oxford).
\bibitem{AKR08} M. Albers, C. Kiefer, and M. Reginatto (2008):
  Measurement analysis and quantum gravity. {\em Phys. Rev. D} {\bf
    78}, 064051.
\bibitem{GRG} C. Kiefer (2009): Quantum geometrodynamics: whence,
  whither? {\em Gen. Relativ. Gravit.} {\bf 41}, 877--901; C. Kiefer
  (2009): Does time exist in quantum gravity? arXiv:0909.3767v1 [gr-qc].
\bibitem{DeWitt}
B. S. DeWitt (1967): Quantum theory of Gravity. I. The canonical
theory. {\em Phys. Rev.} {\bf 160}, 1113--1148.
\bibitem{KMM09} C. Kiefer, J. Marto, and P. V. Moniz (2009):
  Indefinite oscillators and black-hole evaporation. {\em
    Ann. Phys. (Berlin)} {\bf 18}, 722--735. 
\bibitem{HH85} J. J. Halliwell and S. W. Hawking (1985):
Origin of structure in the Universe. {\em Phys. Rev. D} {\bf 31},
1777--1791.
\bibitem{CK87} C. Kiefer (1987): Continuous measurement of
  minisuperspace variables by higher multipoles. {\em Class. Quantum
    Grav.} {\bf 4}, 1369--1382.
\bibitem{Zeh3} H. D. Zeh (1986): Emergence of classical time from a
  universal wave function. {\em Phys. Lett. A} {\bf 116}, 9--12.
\bibitem{deco} E. Joos, H. D. Zeh, C. Kiefer, D. Giulini, J. Kupsch, and
I.-O. Stamatescu (2003):
 {\em Decoherence and the Appearance of a Classical World
in Quantum Theory}, second edition (Springer, Berlin).
\bibitem{Starobinsky} A. A. Starobinsky (1979): Spectrum of relict
  gravitational radiation and the early state of the universe. {\em
    JETP Lett.} {\bf 30}, 682--685.
\bibitem{pointerstates} C. Kiefer, I. Lohmar, D. Polarski, and
  A. A. Starobinsky (2007): Pointer states for primordial fluctuations
  in inflationary cosmology. {\em Class. Quantum Grav.} {\bf 24},
  1699--1718; C. Kiefer and D. Polarski (2009): Why do cosmological
  perturbations look classical to us? {\em Adv. Sci. Lett.} {\bf 2},
  164--173, see also arXiv:0810.0087v2 [astro-ph]. 
\bibitem{verbier} C. Kiefer (2001):  Entropy of gravitational waves
  and primordial fluctuations. In:
     {\em Cosmology and particle physics},
     ed. by J.~Garcia-Bellido, R.~Durrer, and
     M.~Shaposhnikov (American Institute of Physics, New York),
     pp.~499--504.
\bibitem{HMH} R. Holman and L. Mersini-Houghton (2006): Why the
  Universe started from a low entropy state. {\em Phys. Rev. D} {\bf
    74}, 123510.
\bibitem{KZ} C. Kiefer and H. D. Zeh (1995): Arrow of time in a recollapsing
           quantum universe. {\em Phys. Rev. D} {\bf 51}, 4145--4153.
\bibitem{KS09} C. Kiefer and B.~Sandh\"ofer (2009): Quantum
  cosmology. In: {\em Beyond the Big Bang}, ed. by R.~Vaas (Springer,
  Berlin), see also arXiv:0804.0672v2 [gr-qc].
\bibitem{bounces} M. Novello and S. E. P. Bergliaffa (2008): 
 Bouncing cosmologies. {\em Phys. Rep.} {\bf 463}, 127--213.
\bibitem{Bojowald} M. Bojowald (2009): A momentous arrow of time.
Contribution to this volume, see also arXiv:0910.3200v1 [gr-qc].
\bibitem{Hinshaw} G. Hinshaw {\em et al}. (2009):
Five-Year Wilkinson Microwave Anisotropy Probe (WMAP) Observations:
Data Processing, Sky Maps, and Basic Results. 
 {\em Astrophys. J. Suppl.} {\bf 180}, 225--245.
\bibitem{Geyer}  K. H. Geyer (1980): Geometrie der Raum-Zeit der Ma\ss
  bestimmung von Kottler, Weyl und Trefftz. {\em Astronomische Nachrichten}
  {\bf 301}, 135--149.
\end{thebibliography}
\end{document}